\documentclass[10pt,aps,prd,twocolumn,superscriptaddress,floatfix,amsmath,amssymb,amsfonts,longbibliography,nofootinbib]{revtex4-2}

\usepackage[english]{babel}
\usepackage{graphicx}
\usepackage{dcolumn}
\usepackage{bm}

\usepackage{physics}
\usepackage{amsmath,amssymb,mathtools}
\usepackage{booktabs}
\usepackage{tikzsymbols}
\usepackage[utf8]{inputenc}
\usepackage[T1]{fontenc}
\usepackage{array}
\usepackage{booktabs}

\usepackage{tikz}
\usetikzlibrary{arrows.meta, positioning}
\usepackage{lipsum}
\usepackage{hyperref}

\usepackage{pgfplots}
\pgfplotsset{compat=1.18}

\hypersetup{
    colorlinks=true,
   linkcolor=blue,
    filecolor=red,      
    urlcolor=cyan,
}

\begin{document}
\title{\textbf{Can Spontaneous Time-Translation Symmetry Breaking Be Quantized Without Explicit Branches?}}

\author{Partha Nandi}
\email{pnandi@sun.ac.za}

\affiliation{%
National Institute for Theoretical and Computational Sciences (NITheCS),
Stellenbosch 7604, South Africa
}

\affiliation{%
Department of Physics,
Stellenbosch University,
Stellenbosch 7600, South Africa
}


\begin{abstract}
Can classically branched dynamics associated with spontaneous
time-translation-symmetry breaking be quantized without
explicitly retaining the branches of the Hamiltonian? We
investigate this question using a reduced phase-space
formulation in which the Hamiltonian is single-valued, while
the critical structure of the classical Legendre map is
encoded in a singular, momentum-dependent symplectic geometry.
Quantization of the reduced Poisson algebra yields a deformed
position-momentum commutator whose singularities determine
distinct regular momentum sectors. Using the positive measure
induced by the reduced symplectic structure, we identify a
bounded central sector and two unbounded outer sectors. In
the central sector, states with vanishing mean momentum obey
a nonzero lower bound on position uncertainty, subject to the
operator-domain assumptions of the representation. We further
identify a possible field-theoretic connection with $k$-essence,
where nonlinear scalar-field kinetics can produce finite-velocity
degeneracies of the Legendre map. The relationship between
the resulting reduced quantum description and branch-matching
quantization remains unresolved and requires a comparison of
the corresponding Hilbert spaces and operator domains.
\end{abstract}


\maketitle

\smallskip





\section{Introduction}

Symmetry principles lie at the heart of modern physics, providing a powerful framework for understanding the organization of matter and the emergence of collective phenomena across vastly different scales, from cosmological phase transitions~\cite{PhysRevD.39.1558,PhysRevLett.48.1220} to the fundamental interactions governing nuclear and particle physics~\cite{PhysRevLett.13.508}. In many-body quantum systems, spontaneous symmetry breaking (SSB) plays a central role in characterizing equilibrium phases and nonequilibrium dynamical behavior~\cite{Else2016}. A particularly intriguing example is time-crystalline behavior, which raises fundamental questions about the realization of time-translation symmetry breaking in dynamical systems.

A distinctive classical realization of time-crystalline dynamics was proposed by Shapere and Wilczek~\cite{PhysRevLett.109.160402}. In this setting, a non-convex dependence of the Lagrangian on velocity can produce energy minima at nonzero velocities. This mechanism has motivated further investigations of time-crystalline dynamics and related realizations in quantum and spacetime-crystal settings~\cite{Sacha_2018, PhysRevA.91.033617,Zhang:2016kpq,PhysRevLett.120.215301,PhysRevLett.109.163001,PhysRevLett.120.096401}. At the Hamiltonian level, however, non-convexity can obstruct the global invertibility of the canonical momentum--velocity relation, so that the Legendre transformation yields a multivalued Hamiltonian. This raises a basic question about quantization: must the classical branches remain explicit in the quantum description, or can their critical structure be encoded in a reduced phase-space formulation?

The non-invertibility of the Legendre map is central to this question. At its critical points, the momentum--velocity relation ceases to be locally invertible, separating regions in which distinct Hamiltonian branches can be defined. In the branched quantization developed by Shapere and Wilczek~\cite{PhysRevLett.109.200402,Zhao:2012dp}, the branches are retained explicitly, and their matching must be incorporated into the quantum description. Related work has examined the consistency of effective Hamiltonian and Lagrangian descriptions for nonconvex kinetic energies, emphasizing that the validity of the corresponding low-energy descriptions can depend on the coupling to a dissipative environment~\cite{Koliofoti:2025zmy}. Degenerate kinetic structures have also been investigated in cosmological $k$-essence models, where the degeneracy surfaces of the Legendre map can separate dynamical regions and lead to a loss of degrees of freedom~\cite{FerreiraJunior:2023qxi}. These studies underscore the importance of the Legendre map and its degeneracies, but do not settle the operator-domain questions arising in the reduced quantum representation considered here.

In this work, we investigate a complementary approach in which the Hamiltonian is single-valued in reduced variables, while the critical structure of the Legendre map is encoded in a momentum-dependent, noncanonical symplectic structure. Our objective is to determine how this reduced phase-space geometry manifests itself in quantum kinematics, particularly in the commutation relations, the structure of momentum space, and position uncertainty. The approach does not assume that the resulting representation is globally equivalent to explicit branch-matching quantization.

Related investigations have explored connections between time-crystalline behavior, quantum memory, and gravitational phenomena, including gravitationally induced entanglement, memory-driven time-crystalline behavior at finite temperature, and quantum-memory effects in gravitational-radiation-induced entanglement~\cite{Dutta:2025bge,Dutta:2025ouy}. In a different setting, spin-induced phase-space deformation in the Bateman dual oscillator has been shown to generate self-similar temporal scaling and non-Markovian memory~\cite{NandiVitiello2026}. These studies illustrate the broader relevance of phase-space structure to quantum dynamics. The question addressed here is more specific: how does the singular reduced symplectic structure inherited from a non-invertible Legendre map constrain the quantum representation of the reduced system?

Starting from the reduced Poisson algebra $\{y,p_y\}_{\rm SB}=f(p_y)$, we adopt the quantization prescription $[\hat y,\hat p_y]=i\hbar f(\hat p_y)$. Unlike phenomenological generalized uncertainty principle (GUP) models, in which a deformed commutator is postulated at the outset, the deformation considered here is motivated by the reduced classical symplectic structure associated with the Legendre transformation~\cite{Banerjee:2001zi,PhysRevA.36.4417,PhysRevD.90.024003}. Momentum-dependent deformations of the Heisenberg algebra have also been studied in approaches motivated by quantum gravity, string theory, and noncommutative geometry~\cite{Bosso:2023aht,Pramanik:2013zy,PhysRevD.104.126010,Dabrowski:2020ixn,Wagner:2022bcy,Mignemi2003,SW4,SW5,GROSS1988407,SW6,SW6+,Gangopadhyay2019,Bosso:2023sxr}. Such deformations have been investigated in quantum-optical systems, including proposals to constrain Planck-scale effects through gravitational-wave interferometry~\cite{Bosso:2018ckz}. Bosso has also examined momentum-space representations in minimal-length quantum mechanics, discussing the relation between the deformation function, the momentum-space measure, and restrictions to sign-definite momentum domains~\cite{Bosso:2020aqm}. Singular momentum-dependent deformations have been studied by Pedram, who considered the algebra $[X,P]=i\hbar/(1-\beta P^2)$, which becomes singular at finite momentum~\cite{Pedram:2011gw}. Related nonlinear deformations have also been investigated in finite-dimensional quantum registers, where the analytic structure of the deformation determines the associated Walsh--Pauli spectrum~\cite{DeSousa:2026ziz}.

The distinguishing feature of the present construction is that the deformation function is determined by the reduced classical phase-space geometry, with its singularities inherited from the critical points of the Legendre map. The commutator above specifies our quantization prescription for the reduced algebra; it does not establish the uniqueness of the quantum representation. In particular, the choice of Hilbert space, inner product, operator domains, and boundary conditions must be examined independently.

The zeros of the reduced symplectic Jacobian divide the momentum axis into five regular intervals. We focus on a positive-inner-product representation defined on the three selected intervals where $D(p_y)>0$, using the measure induced by the reduced symplectic structure. This is a choice of representation, not a claim that the complementary intervals are inadmissible under every possible quantization prescription. Introducing the natural momentum coordinate $q=F(p_y)$, we express the formal position operator as $i\hbar\,d/dq$ and investigate its operator-domain properties. The central sector admits self-adjoint extensions of the position operator, whereas the derivative operator on either individual outer half-line has no self-adjoint extension on that half-line Hilbert space alone. In the central sector, the deformed commutator also yields a conditional lower bound on position uncertainty for normalized states with vanishing mean momentum and support strictly inside the bounded momentum interval. This result is conditional on the selected representation and the stated domain and state assumptions; it does not establish a universal minimal length for the full theory.

We also discuss a possible connection with nonlinear scalar-field dynamics in cosmology. In particular, $k$-essence models can exhibit finite-velocity degeneracies of the homogeneous Legendre map for suitable kinetic functions~\cite{FerreiraJunior:2023qxi}. This provides a structural analogy with the degeneracy mechanism considered here, but does not by itself establish that the same reduced symplectic structure emerges from a complete gravitational reduction. Establishing such a connection would require a separate constrained Hamiltonian analysis.

The contribution of this work is a sectorwise analysis of the quantum kinematics induced by the reduced symplectic structure of a classically branched system. We construct a positive-inner-product representation on selected regular momentum intervals, examine the self-adjointness properties of the position operator on those sectors, and derive a conditional localization bound in the bounded central sector. We also identify a local correspondence between the reduced and branchwise differential expressions in the natural momentum coordinate. This correspondence does not establish global quantum equivalence: a complete comparison requires compatible Hilbert spaces, operator domains, boundary or matching conditions, and physical observables. The relation between the reduced representation and explicit branch-matching quantization therefore remains an open problem.

The paper is organized as follows. In Sec.~II, we introduce a stabilized non-convex velocity-dependent model and construct its single-valued reduced Hamiltonian and noncanonical symplectic structure. We identify the critical points of the momentum--velocity relation and the finite-velocity ground states of the model. In Sec.~III, we quantize the reduced algebra, construct the momentum-space representation, and analyze its regular sectors, operator domains, and uncertainty structure. Section~IV discusses the possible field-theoretic connection and summarizes the results and outstanding questions.

\section{Spontaneous Breaking of Time-Translation Symmetry and Reduced Noncanonical Phase Space}

We begin by considering a class of higher-order velocity-dependent
systems whose equations of motion contain no more than two time
derivatives. Such systems provide a natural setting in which a
non-convex kinetic sector can coexist with ordinary low-velocity
dynamics and can give rise to spontaneous breaking of continuous
time-translation symmetry. We restrict attention to Lagrangians that
depend on the configuration variable $y$ and its first time derivative
$\dot y$, but contain no higher time derivatives. A general polynomial
kinetic sector can then be written as
\begin{equation}
\label{generalL}
L_N
=
\sum_{n=1}^{N}
\frac{1}{2n}f_n(y)\dot y^{2n}
-
V(y),
\qquad N>1,
\end{equation}
where $f_n(y)$ are functions of $y$. We assume that the highest-order
coefficient satisfies $f_N>0$, so that the kinetic energy is bounded
from below at sufficiently large velocity, while at least one of the
lower-order coefficients is negative. These conditions allow the
kinetic sector to become non-convex and hence to develop stationary
points at finite velocity.

The essential feature is the non-monotonicity of the relation between
the canonical momentum and the generalized velocity,
\begin{equation}
p_y^c
=
\frac{\partial L_N}{\partial\dot y},
\end{equation}
with $\frac{\partial p_y^c}{\partial\dot y}=0$ at finite values of
$\dot y$. Consequently, the map $\dot y\mapsto p_y^c$ ceases to be
globally invertible, and the Hamiltonian expressed in terms of the
canonical momentum is generally multivalued. This is the characteristic
branch structure associated with the Shapere--Wilczek class of
time-translation-symmetry-breaking systems~\cite{PhysRevLett.109.160402}.

The class of non-convex velocity-dependent systems considered here is
not restricted to formal mechanical models. Effective
velocity-dependent kinetic structures of this type arise in physically
motivated settings, including the Shapere--Wilczek mechanism and
related nonlinear systems. In particular, a minisuperspace reduction
of quadratic-curvature gravity has been shown to produce an effective
Shapere--Wilczek-type velocity-dependent model with a nonzero-velocity
ground state and a noncanonical phase-space structure
\cite{PhysRevD.98.024004}. These examples motivate the study of the reduced
phase-space geometry associated with non-convex kinetic dynamics.

\subsection{A Representative Stabilized Model}

To make the quantum consequences of the reduced phase-space
description explicit, we consider a concrete representative of the
class of velocity-dependent systems introduced above. The purpose of
this model is not to propose a unique microscopic realization, but to
provide the simplest stabilized effective model that simultaneously
captures the three ingredients relevant for our analysis:
(i) conventional quadratic behavior at low velocity,
(ii) non-convexity over an intermediate velocity range, and
(iii) stabilization of the kinetic sector at large velocity.

The minimal even polynomial with these properties is
\begin{equation}
T(\dot y)
=
\frac{1}{2}m\dot y^2
-\frac{\alpha}{4}\dot y^4
+\frac{\gamma}{6}\dot y^6,
\qquad
\alpha>0,\quad \gamma>0 .
\label{genericT}
\end{equation}
The quadratic term guarantees the usual low-velocity kinetic behavior,
while the negative quartic term produces the non-convexity responsible
for the non-monotonic momentum--velocity relation. The positive sextic
term stabilizes the kinetic energy at large velocity. Importantly, the
sextic term is not required for the existence of the branched structure
itself; it provides a controlled stabilized representative in which
the reduced phase-space geometry can be analyzed analytically.

For convenience, we parameterize the higher-order coefficients as
\begin{equation}
\alpha=\beta m^3,
\qquad
\gamma=\beta'^2m^5 ,
\end{equation}
and consider
\begin{equation}
L_3
=
\frac{1}{2}m\dot y^2
-\frac{1}{4}\beta m^3\dot y^4
+\frac{1}{6}\beta'^2m^5\dot y^6
-
V(y).
\label{12}
\end{equation}
The parameters $\beta$ and $\beta'$ characterize the effective
higher-order velocity corrections. The particular parameter relation
used below is only illustrative and is not essential to the
construction.

The canonical momentum is
\begin{equation}
p_y^c
=
\frac{\partial L_3}{\partial\dot y}
=
m\dot y-\beta m^3\dot y^3
+\beta'^2m^5\dot y^5 .
\label{canonicalmomentum}
\end{equation}
The crucial feature is that this relation is not globally monotonic.
Its local invertibility is determined by
\begin{equation}
\frac{\partial p_y^c}{\partial\dot y}
=
m\left(
1-3\beta m^2\dot y^2
+5\beta'^2m^4\dot y^4
\right).
\label{differentialmass}
\end{equation}
The zeros of this derivative are therefore the turning points of the
momentum--velocity map. At these points the Legendre transformation
ceases to be locally invertible. Consequently, a fixed canonical
momentum can correspond to several values of the velocity and hence to
several values of the Hamiltonian. This is the origin of the branched
Hamiltonian structure.

\begin{figure*}[t]
    \centering
    \includegraphics[width=0.95\textwidth]{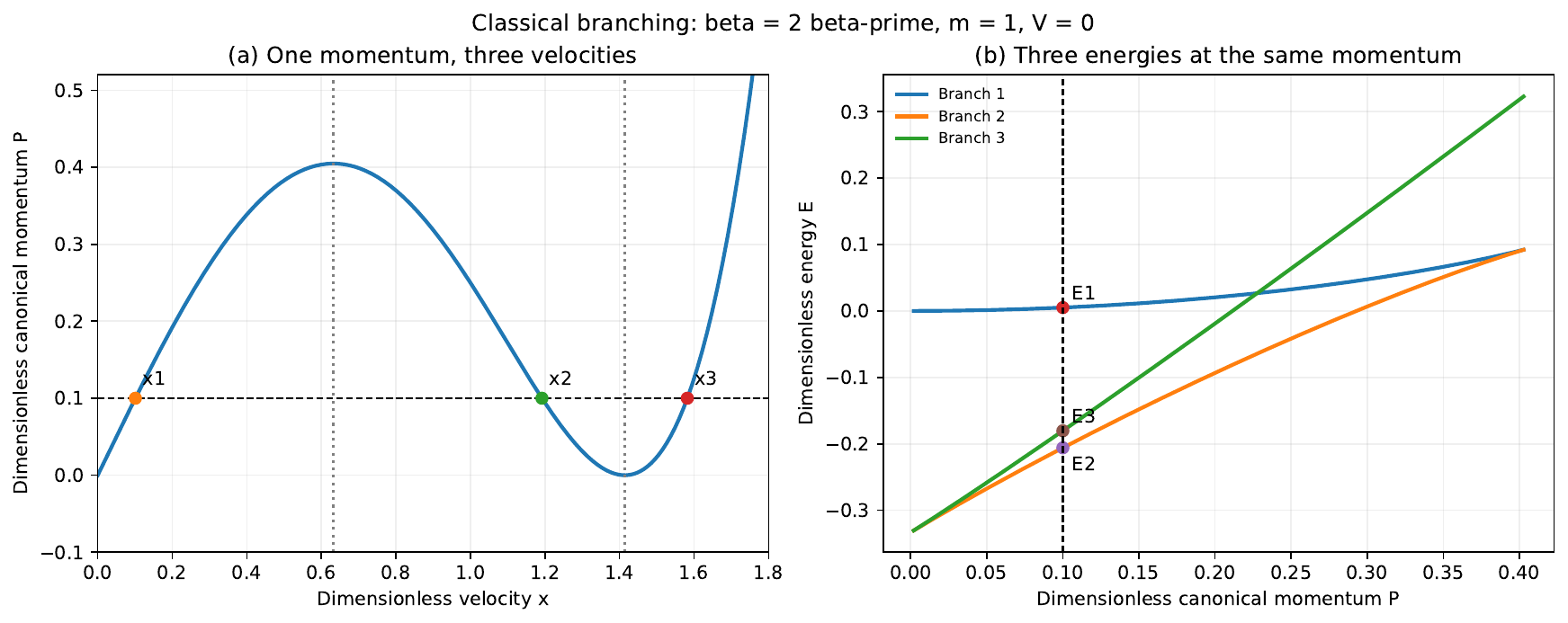}
    \caption{Origin of the branched Hamiltonian structure for the
    representative choice $\beta=2\beta'$. In panel (a), a fixed
    canonical momentum $p_y^c$ intersects the non-monotonic
    momentum--velocity relation at three distinct velocities
    $\dot y_1,\dot y_2,\dot y_3$. Panel (b) shows the corresponding
    three Hamiltonian values at the same canonical momentum. The
    figure illustrates how the non-invertibility of the Legendre
    transformation gives rise to the branched Hamiltonian structure.}
    \label{fig:branched_structure}
\end{figure*}

This structure is illustrated schematically in
Fig.~\ref{fig:branched_structure}. In particular, for a value of
$p_y^c$ lying in the region where the momentum--velocity map is
multivalued, three distinct velocities,
$\dot y_1$, $\dot y_2$, and $\dot y_3$, are possible. The corresponding
Legendre transforms give three values of the Hamiltonian,
$H_1$, $H_2$, and $H_3$, at the same canonical momentum. Thus the
multivaluedness is not an additional assumption: it follows directly
from the non-invertibility of the canonical momentum--velocity
relation.

The multivalued canonical Hamiltonian is precisely the structure that
makes a conventional global canonical quantization nontrivial. Rather
than retaining the different Hamiltonian branches as independent
quantum components, as in the branched quantization approach, we now
ask whether the same classical dynamics can be represented by a
single-valued Hamiltonian on a suitably chosen reduced phase space.

To this end, we introduce the mechanical momentum
\begin{equation}
p_y=m\dot y .
\label{mechanicalmomentum}
\end{equation}
This variable should not be interpreted as a canonical transformation
of $p_y^c$. Instead, it provides a convenient coordinate on the reduced
phase space. In terms of this variable,
\begin{equation}
p_y^c
=
p_y-\beta p_y^3+\beta'^2p_y^5 .
\end{equation}
Thus the nonlinear relation responsible for the original branching is
still present; it has simply been expressed as a nonlinear relation
between the canonical and reduced momenta.

Using $p_y=m\dot y$, the Legendre expression for the energy can be
written as the single-valued function
\begin{equation}
H_c
=
\frac{p_y^2}{2m}
-\frac{3\beta}{4m}p_y^4
+\frac{5\beta'^2}{6m}p_y^6
+
V(y).
\label{c}
\end{equation}
This is the key classical reformulation used in the remainder of the
paper. The branching has not been eliminated dynamically. Rather, the
Hamiltonian is now single-valued, while the information associated
with the non-invertible Legendre map is transferred to the symplectic
structure of the reduced phase space.

The corresponding first-order reduced Lagrangian is
\begin{equation}
\begin{aligned}
L_{3f}
={}&
\dot y\,p_y
-\beta\dot y\,p_y^3
+\beta'^2\dot y\,p_y^5
-\frac{p_y^2}{2m}
+\frac{3\beta}{4m}p_y^4
\\
&\hspace{2cm}
-\frac{5\beta'^2}{6m}p_y^6
-V(y).
\end{aligned}
\label{reducedfirstorder}
\end{equation}

Variation with respect to $y$ and $p_y$ gives
\begin{equation}
\left(
1-3\beta p_y^2+5\beta'^2p_y^4
\right)\dot p_y
=
-\frac{\partial H_c}{\partial y},
\label{eom1}
\end{equation}
and
\begin{equation}
\left(
1-3\beta p_y^2+5\beta'^2p_y^4
\right)\dot y
=
\frac{\partial H_c}{\partial p_y}.
\label{eom2}
\end{equation}

Introducing the reduced phase-space coordinates
\begin{equation}
\xi_i=(y,p_y),
\qquad i=1,2,
\end{equation}
these equations take the symplectic form \cite{Wr}
\begin{equation}
\Lambda'_{ij}\dot{\xi}_j
=
\frac{\partial H_c}{\partial\xi_i},
\label{symplecticeq}
\end{equation}
with
\begin{equation}
\Lambda'_{ij}
=
\begin{pmatrix}
0 & -\mathcal{D}(p_y)\\
\mathcal{D}(p_y) & 0
\end{pmatrix},
\end{equation}
where
\begin{equation}
\mathcal{D}(p_y)
=
1-3\beta p_y^2+5\beta'^2p_y^4 .
\label{Ddef}
\end{equation}

Away from the zeros of $\mathcal{D}(p_y)$, the inverse symplectic
matrix is
\begin{equation}
(\Lambda'^{-1})^{ij}
=
\begin{pmatrix}
0 & \mathcal{D}(p_y)^{-1}\\
-\mathcal{D}(p_y)^{-1} & 0
\end{pmatrix}.
\end{equation}
The corresponding reduced Poisson bracket is therefore
\begin{equation}
\{A,B\}_{\rm SB}
=
(\Lambda'^{-1})^{ij}
\frac{\partial A}{\partial\xi_i}
\frac{\partial B}{\partial\xi_j},
\end{equation}
so that
\begin{equation}
\{y,p_y\}_{\rm SB}=f(p_y),
\end{equation}
with
\begin{equation}
f(p_y)
=
\frac{1}
{1-3\beta p_y^2+5\beta'^2p_y^4}.
\label{fclassical}
\end{equation}

The important point is therefore that the multivaluedness of the
original canonical Hamiltonian has been transferred to the reduced
phase-space geometry. The reduced Hamiltonian is single-valued, but
the reduced Poisson bracket becomes singular precisely where the
original momentum--velocity relation ceases to be locally invertible.
Thus, the reduced description does not eliminate the classical
branch structure; it encodes its critical information geometrically.

The critical locus is determined by
\begin{equation}
\mathcal{D}(p_y)=0,
\end{equation}
or equivalently
\begin{equation}
1-3\beta p_y^2+5\beta'^2p_y^4=0.
\end{equation}
At these points the reduced symplectic form becomes degenerate and
the reduced Poisson bracket is singular. The singular locus therefore
separates regions in which the reduced phase-space description is
regular. This geometric sector structure will be inherited by the
quantum representation developed in Sec.~III.

\subsection{Ground States and Spontaneous Breaking of
Time-Translation Symmetry}

We now determine the minimum-energy configurations of the reduced
Hamiltonian and examine their relation to the singular structure of
the reduced phase space. For a spatially constant potential, the
stationary points of the energy landscape satisfy
\begin{equation}
\frac{\partial H_c}{\partial p_y}
=
\frac{p_y}{m}
\left(
1-3\beta p_y^2+5\beta'^2p_y^4
\right)
=0.
\end{equation}
Besides the stationary point at $p_y=0$, the nonzero stationary
points are determined by
\begin{equation}
1-3\beta p_y^2+5\beta'^2p_y^4=0.
\end{equation}
Assuming $\beta>0$ and $\beta'>0$, the corresponding positive
critical momentum magnitudes $p_-<p_+$ are given by
\begin{equation}
\label{kj}
p_\pm^2
=
\frac{3\beta}{10\beta'^2}
\left[
1\pm
\sqrt{
1-\frac{20}{9}
\left(\frac{\beta'}{\beta}\right)^2
}
\right].
\end{equation}
Two distinct positive critical values of $p_y^2$ exist provided
\begin{equation}
\frac{\beta}{\beta'}>\frac{2\sqrt{5}}{3}.
\end{equation}
At equality, the two critical values coalesce, whereas below
this threshold no real nonzero stationary points exist.

The second derivative of $H_c$ shows that the inner stationary
points $\pm p_-$ are local maxima, whereas the outer stationary
points $\pm p_+$ are local minima. The nonzero minimum-energy
configurations are consequently
\begin{equation}
p_{y,\mathrm{gs}}=\pm p_+ .
\end{equation}
Since the reduced momentum is related to the velocity by
$p_y=m\dot y$, these minima correspond to
\begin{equation}
\dot y_{\mathrm{gs}}
=
\pm\frac{p_+}{m}.
\end{equation}
For a constant potential, the corresponding trajectories are
\begin{equation}
y_{\mathrm{gs}}(t)
=
y_0+\frac{p_{y,\mathrm{gs}}}{m}t .
\end{equation}
Thus, the minimum-energy configurations are not static but exhibit
persistent motion with nonzero velocity. In this sense, the model
realizes spontaneous breaking of continuous time-translation
symmetry: the lowest-energy configurations are dynamically moving
rather than time-independent. The two signs of
$p_{y,\mathrm{gs}}$ correspond to the two degenerate families of
ground-state motion.

\begin{figure}[t]
    \centering
    \includegraphics[width=0.85\linewidth]{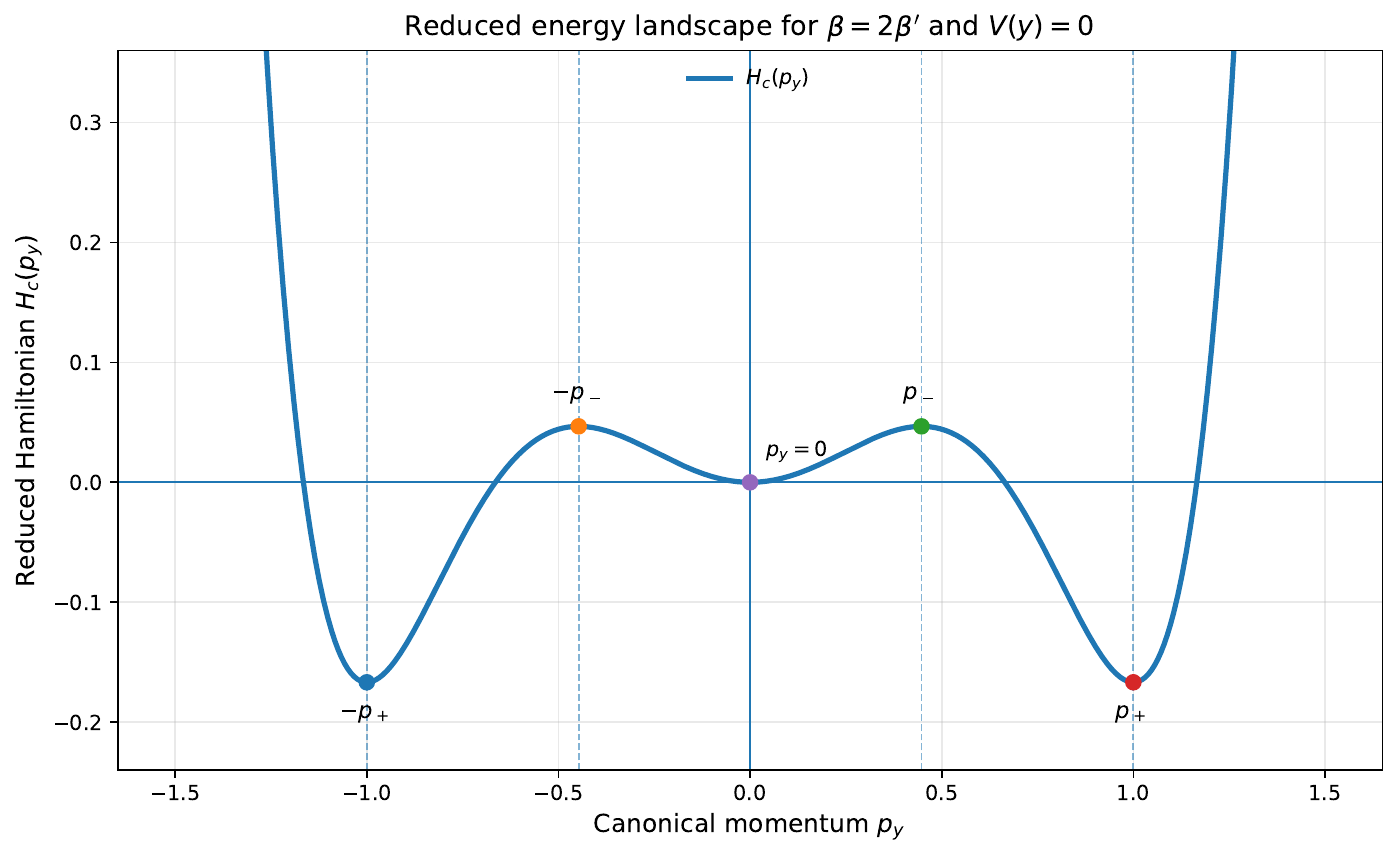}
    \caption{Reduced energy $H_c(p_y)$ for the representative choice
$\beta=2\beta'$, with $V(y)=0$. The outer minima at
$p_y=\pm p_+$ correspond to the finite-velocity minimum-energy
configurations, while the inner stationary points
$p_y=\pm p_-$ are local maxima, where
$p_-^2=2/(5\beta)$ and $p_+^2=2/\beta$.}
    \label{fig:energy_landscape}
\end{figure}

For the representative choice
\begin{equation}
\beta=2\beta',
\end{equation}
the roots simplify to
\begin{equation}
\label{specialroots}
p_-^2=\frac{2}{5\beta},
\qquad
p_+^2=\frac{2}{\beta}.
\end{equation}
The corresponding energy landscape is shown in
Fig.~\ref{fig:energy_landscape}. The outer minima at
$p_y=\pm p_+$ represent the finite-velocity ground states, whereas
the inner stationary points $\pm p_-$ are local maxima.

An important feature of the reduced description is that the nonzero
stationary points of the energy landscape coincide with the singular
locus of the reduced symplectic structure. This connection is encoded
in the factor
\begin{equation}
\mathcal{D}(p_y)
=
1-3\beta p_y^2+5\beta'^2p_y^4,
\end{equation}
which governs the degeneracy of the reduced symplectic form. The
four nonzero singular points are
\begin{equation}
p_y=\pm p_+,\qquad p_y=\pm p_-,
\end{equation}
where $p_+>p_->0$ are the positive roots determined by
\begin{equation}
\mathcal{D}(\pm p_\pm)=0.
\end{equation}
The outer singular points $\pm p_+$ therefore coincide with the
finite-momentum ground states, whereas the inner singular points
$\pm p_-$ correspond to the intervening energy maxima.

This coincidence provides the central geometric link between the
classical dynamics and the reduced phase-space description. The
critical points of the momentum--velocity relation, which are
responsible for the failure of the canonical Legendre transformation,
are precisely the points at which the reduced symplectic structure
becomes degenerate. Among these critical points, the outer pair
selects the minimum-energy configurations with nonzero velocity.

At the singular points, the reduced Hamilton equation for $\dot y$
takes the indeterminate form $0/0$. The ground-state velocity should
therefore not be inferred by evaluating that equation directly at
the degenerate point. Instead, it follows from the defining relation
between the reduced momentum and velocity,
\begin{equation}
\dot y_{\mathrm{gs}}
=
\pm\frac{p_+}{m}.
\end{equation}

The same critical momenta subsequently appear as singularities of
the momentum-dependent quantum algebra. Thus, the classical
finite-velocity ground states and the boundaries of the quantum
momentum sectors are controlled by the same critical structure.

This structure also has a simple global interpretation. The singular
locus of the reduced phase space is
\begin{equation}
\Sigma
=
\left\{
(y,p_y)\,:\,
\mathcal{D}(p_y)=0
\right\},
\end{equation}
and the corresponding regular phase space is
\begin{equation}
\Gamma_{\rm reg}
=
\Gamma_{\rm red}\setminus\Sigma .
\end{equation}
The critical points $p_y=\pm p_\pm$ therefore separate the regular
reduced phase space into disconnected components. In momentum space,
these singular points act as boundaries between regions in which the
reduced symplectic structure is nonsingular. The momentum-sector
decomposition that emerges after quantization is consequently not an
arbitrary partition of momentum space; it is inherited from the
connected-component structure of the regular reduced phase space.

This observation provides a geometric bridge between the classical
branch structure and the quantum sector structure. In the original
canonical description, the critical points arise from the
non-invertibility of the Legendre transformation and are associated
with the meeting of Hamiltonian branches. In the reduced description,
the Hamiltonian is single-valued, but the same critical structure is
encoded in the singular symplectic geometry. After quantization, these
singular points become boundaries of the regular momentum
representation. We therefore regard the resulting quantum sectors as
geometric descendants of the classical branch structure. We do not
interpret them as distinct topological phases, since no nontrivial
topological invariant is constructed here.

\subsection{A Possible Field-Theoretic Connection: \texorpdfstring{$k$-Essence}{k-Essence}}
\label{subsec:kessence}

A possible field-theoretic setting for investigating nonlinear
kinetic dynamics is provided by $k$-essence, in which the
scalar-field Lagrangian depends nonlinearly on its kinetic
invariant~\cite{ArmendarizPiconMukhanovSteinhardt2001}.
The purpose of this subsection is to investigate whether a
homogeneous scalar field can exhibit a Legendre-map degeneracy
analogous to that encountered in the mechanical model studied
above.

We consider the scalar--gravity action
\begin{equation}
S=\int d^4x\,\sqrt{-g}\left[
\frac{M_{\mathrm{Pl}}^2}{2}R+P(X)
\right],
\label{eq:kessence_action}
\end{equation}
where $R$ is the Ricci scalar, $M_{\mathrm{Pl}}$ is the
reduced Planck mass, and $P(X)$ is the scalar-field Lagrangian
density. The kinetic invariant is
\begin{equation}
X=-\frac{1}{2}g^{\mu\nu}
\partial_\mu\phi\,\partial_\nu\phi,
\label{eq:kessence_X}
\end{equation}
where $\phi$ denotes the scalar field. We use natural units
$c=\hbar=1$ and the conventional four-dimensional
normalization $[\phi]=\mathrm{mass}$.

As an illustrative example, consider the polynomial kinetic
function
\begin{equation}
P(X)=X+\lambda X^2+\mu X^3,
\label{eq:kessence_polynomial}
\end{equation}
where $\lambda$ and $\mu$ are higher-order kinetic couplings
with mass dimensions $[\lambda]=\mathrm{mass}^{-4}$ and
$[\mu]=\mathrm{mass}^{-8}$, respectively.

To examine the homogeneous dynamics, we consider a spatially
flat Friedmann--Lemaître--Robertson--Walker (FLRW) spacetime,
\begin{equation}
ds^2=-N^2(t)\,dt^2+a^2(t)\,d\mathbf{x}^2,
\label{eq:kessence_FLRW}
\end{equation}
for which
\begin{equation}
X=\frac{\dot{\phi}^{\,2}}{2N^2}.
\label{eq:kessence_homogeneous_X}
\end{equation}
Here $N(t)$ is the lapse function, $a(t)$ is the scale
factor, and an overdot denotes differentiation with respect
to the coordinate time $t$. Defining the proper-time
velocity $v=\dot{\phi}/N$ and taking the fiducial comoving
spatial volume to be unity, the scalar contribution to the
minisuperspace Lagrangian is
\begin{equation}
L_\phi=a^3N\left(
\frac{v^2}{2}
+\frac{\lambda v^4}{4}
+\frac{\mu v^6}{8}
\right).
\label{eq:kessence_homogeneous_L}
\end{equation}

The momentum conjugate to $\phi$ is
\begin{equation}
\pi_\phi
:=\frac{\partial L_\phi}{\partial\dot{\phi}}
=a^3\left(
v+\lambda v^3+\frac{3\mu}{4}v^5
\right).
\label{eq:kessence_momentum}
\end{equation}
The local invertibility of the homogeneous Legendre map
$\dot{\phi}\mapsto\pi_\phi$ is controlled by its Jacobian,
namely the derivative of the momentum with respect to the
coordinate-time velocity:
\begin{equation}
\frac{\partial\pi_\phi}{\partial\dot{\phi}}
=\frac{a^3}{N}\left(
1+3\lambda v^2+\frac{15\mu}{4}v^4
\right).
\label{eq:kessence_jacobian}
\end{equation}
For $a>0$ and $N\ne0$, the Legendre map becomes locally
singular when this Jacobian vanishes, or equivalently when
\begin{equation}
1+3\lambda v^2+\frac{15\mu}{4}v^4=0.
\label{eq:kessence_degeneracy}
\end{equation}
For $\lambda<0$ and $\mu>0$, two distinct positive roots
for $v^2$ exist when $9\lambda^2>15\mu$. The corresponding
critical velocities satisfy
\begin{equation}
v_\pm^2
=\frac{2}{5\mu}
\left(
-\lambda\pm\sqrt{\lambda^2-\frac{5\mu}{3}}
\right).
\label{eq:kessence_critical_velocities}
\end{equation}
These roots identify finite-velocity degeneracies of the
homogeneous scalar Legendre map.

The relevance to the mechanical model is structural:
in both systems, local invertibility of the momentum--velocity
relation fails when the corresponding Legendre-map Jacobian
vanishes. The scalar-field condition derived here does not,
however, establish the same reduced Poisson bracket as that
obtained in the mechanical model. In particular, the existence
of critical velocities in the homogeneous scalar sector does
not demonstrate that the same reduced phase-space structure
emerges after imposing the gravitational constraints.

A further issue concerns the propagation and stability of
scalar perturbations. For $k$-essence, the usual expression
for the squared sound speed is
\begin{equation}
c_s^2=\frac{P_X}{P_X+2XP_{XX}},
\qquad
P_X:=\frac{\partial P}{\partial X},
\qquad
P_{XX}:=\frac{\partial^2P}{\partial X^2}.
\label{eq:kessence_sound_speed}
\end{equation}
For the polynomial in Eq.~(\ref{eq:kessence_polynomial}),
the denominator becomes
\begin{equation}
P_X+2XP_{XX}
=1+6\lambda X+15\mu X^2.
\label{eq:kessence_kinetic_coefficient}
\end{equation}
Using $X=v^2/2$, this expression coincides with the
velocity-dependent factor in the Legendre-map Jacobian
in Eq.~(\ref{eq:kessence_jacobian}), apart from the
nonvanishing prefactor $a^3/N$. Thus, in this homogeneous
model, the same degeneracy condition appears in the
Legendre map and in the scalar kinetic coefficient relevant
to perturbations. The vanishing of this coefficient requires
a separate analysis of the perturbation dynamics and does
not, by itself, establish the stability or instability of
a cosmological background.

We therefore regard this example as a possible
field-theoretic setting in which to investigate nonlinear
Legendre-map degeneracies, rather than as an established
gravitational realization of the reduced phase-space
structure studied here. Establishing a stronger connection
requires a constrained Hamiltonian analysis of the coupled
gravity--scalar system, including the lapse constraint,
gauge fixing, and reduction to the physical phase space.
Such an analysis would determine whether the homogeneous
degeneracy survives in the physical degrees of freedom
and whether it induces a reduced symplectic structure
analogous to that of the mechanical model.

With this possible field-theoretic motivation in mind,
we now return to the reduced phase-space structure of the
mechanical model and investigate its implications for
quantum kinematics.

\section{Quantum Implications of the SSB Geometry:
Deformed Algebra and Uncertainty Bounds}
\label{sec:quantum}

The reduced classical phase space obtained in Sec.~II is
characterized by the momentum-dependent Poisson bracket
\begin{equation}
    \{y,p_y\}_{\rm SB}=f(p_y),
\end{equation}
where
\begin{equation}
    \label{fdef}
    f(p_y)
    =
    \left(
    1-3\beta p_y^2+5\beta'^2p_y^4
    \right)^{-1}.
\end{equation}
The central question is whether the nontrivial phase-space
geometry associated with the classical branched dynamics
leaves a corresponding imprint on quantum kinematics.
We address this question by quantizing the reduced algebra
and examining its implications for the quantum commutator,
momentum-space representation, Hilbert-space structure, and
position--momentum uncertainty relation.

Momentum-dependent deformations of the Heisenberg algebra
are familiar from generalized-uncertainty-principle (GUP)
frameworks, where modified quantum commutators are related
to deformations of classical Poisson and symplectic structures
\cite{Bruno:2024mss}. The present construction has a different
starting point. Rather than postulating a deformed quantum
algebra and subsequently seeking a classical realization,
we take the deformation to be inherited from the reduced
symplectic structure of the classical branched dynamics.

We therefore adopt the quantization prescription
\begin{equation}
    \label{def}
    [\hat y,\hat p_y]
    =
    i\hbar f(\hat p_y).
\end{equation}
The resulting commutator has the momentum-dependent form
characteristic of GUP-type deformations~\cite{Bruno:2024mss}.
Here, however, $f(p_y)$ is fixed by the reduced classical
phase-space structure rather than introduced as an independent
phenomenological modification. The uncertainty relation derived
below is consequently a formal quantum implication of the
chosen reduced algebra, subject to the operator-domain
conditions discussed in this section.

Equation~\eqref{def} is the quantization prescription adopted
in this work; we do not claim that it is the unique possible
quantization of the reduced classical algebra. In particular,
a complete comparison with the branched quantization of
Shapere and Wilczek~\cite{PhysRevLett.109.200402} requires an analysis of the
corresponding Hilbert spaces, operator domains, boundary
conditions, and physical observables.

A momentum-space representation of Eq.~\eqref{def} is
formally given by
\begin{equation}
    \label{operators}
    \hat p_y=p_y,
    \qquad
    \hat y=i\hbar f(p_y)\frac{\partial}{\partial p_y}.
\end{equation}
Acting on a differentiable test function $\phi(p_y)$ supported
inside a regular momentum interval gives
\begin{equation}
    [\hat y,\hat p_y]\phi(p_y)
    =
    i\hbar f(p_y)\phi(p_y),
\end{equation}
and hence reproduces Eq.~\eqref{def} on this test-function
domain. This formal representation does not, by itself,
establish a self-adjoint realization of $\hat y$ or determine
the global operator domains.

\begin{figure}[t]
    \centering
    \includegraphics[width=0.85\linewidth]
    {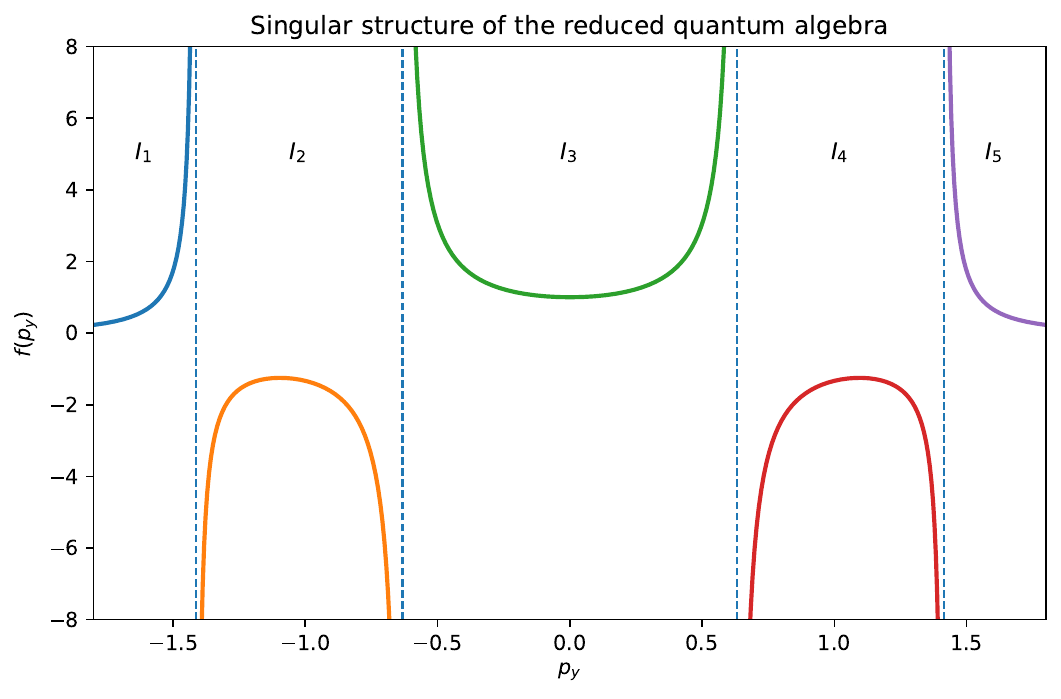}
    \caption{Momentum dependence of the reduced deformation
    function $f(p_y)$ for the representative choice
    $\beta=2\beta'$. The divergences at $p_y=\pm p_\pm$
    occur at the critical points of the classical
    momentum--velocity relation, where the reduced symplectic
    structure becomes degenerate. The outer singularities
    $\pm p_+$ coincide with the finite-momentum classical
    ground-state configurations associated with spontaneous
    time-translation symmetry breaking.}
    \label{fig:deformation_function}
\end{figure}

\subsection{Singularities and Quantum Momentum Sectors}
\label{subsec:momentum_sectors}

The reduced Poisson structure leads to the deformed commutation
relation
\begin{equation}
    [\hat y,\hat p_y]
    =
    i\hbar f(\hat p_y),
    \qquad
    f(p_y)=\frac{1}{D(p_y)},
    \label{eq:deformed_commutator}
\end{equation}
where
\begin{equation}
    D(p_y)
    =
    1-3\beta p_y^2+5\beta'^2p_y^4.
    \label{eq:momentum_denominator}
\end{equation}
The function $f(p_y)$ is singular at the zeros of $D(p_y)$.
For $\beta,\beta'>0$ and
$\beta/\beta'>2\sqrt{5}/3$, the two distinct positive
critical momenta are
\begin{equation}
    p_\pm^2
    =
    \frac{3\beta}{10\beta'^2}
    \left(
    1\pm\sqrt{1-\frac{20\beta'^2}{9\beta^2}}
    \right),
    \label{eq:critical_momenta}
\end{equation}
with $0<p_-<p_+$. The four singular points
$p_y=\pm p_-$ and $p_y=\pm p_+$ divide the real momentum
axis into five regular intervals:
\begin{align}
    I_1&=(-\infty,-p_+),&
    I_2&=(-p_+,-p_-),\nonumber\\
    I_3&=(-p_-,p_-),&
    I_4&=(p_-,p_+),\nonumber\\
    I_5&=(p_+,\infty).
    \label{eq:five_momentum_intervals}
\end{align}
The function $f(p_y)$ is positive on $I_1$, $I_3$, and
$I_5$, and negative on $I_2$ and $I_4$.

\paragraph{Choice of momentum sectors.}

We construct the quantum representation on the selected
regular domain
\begin{equation}
    \mathcal P_+
    =
    I_1\cup I_3\cup I_5
    =
    (-\infty,-p_+)\cup(-p_-,p_-)\cup(p_+,\infty).
    \label{eq:chosen_momentum_sectors}
\end{equation}
The choice is motivated by the momentum-space measure
adopted from the reduced symplectic structure,
\begin{equation}
    d\mu(p_y)
    =
    \frac{dp_y}{f(p_y)}
    =
    D(p_y)\,dp_y.
    \label{eq:positive_momentum_measure}
\end{equation}
The relationship between the deformation function, the
momentum-space measure, and restrictions to sign-definite
momentum domains has also been discussed in the minimal-length
literature~\cite{Bosso:2020aqm}. On $\mathcal P_+$, the
measure in Eq.~\eqref{eq:positive_momentum_measure} is
positive and therefore defines a positive inner product for
the representation considered here. Moreover, $f(p_y)$
has the same sign as in the canonical commutation relation
on these sectors.

The significance of the sign can be seen by considering
the quadratic form associated with the signed measure
$D(p_y)\,dp_y$:
\begin{equation}
    \|\psi\|_D^2
    =
    \int_{\mathbb R}
    D(p_y)|\psi(p_y)|^2\,dp_y.
    \label{eq:signed_measure_norm}
\end{equation}
If this signed measure were used on a state space admitting
arbitrary smooth wavefunctions supported in all five
intervals, any nonzero wavefunction with compact support
strictly inside $I_2$ or $I_4$ would have negative norm.
Thus, the signed measure does not define a positive-definite
inner product on that state space. This motivates restricting
the present construction to $\mathcal P_+$, but does not
establish that the other regular intervals cannot be
quantized using a different prescription.

For example, on an interval where $f(p_y)<0$, one could
consider the positive density
\begin{equation}
    d\mu_{\mathrm{alt}}(p_y)
    =
    \frac{dp_y}{|f(p_y)|}
    =
    |D(p_y)|\,dp_y.
    \label{eq:alternative_positive_measure}
\end{equation}
This density is not the signed symplectic volume element
adopted in Eq.~\eqref{eq:positive_momentum_measure};
it represents an alternative choice of positive measure.
Importantly, adopting it would not, by itself, require
replacing $f(p_y)$ by $|f(p_y)|$ in the commutation
relation. The original deformation function can be retained
in the formal differential representation, and the
position operator is locally symmetric on compactly
supported test functions within an individual regular
interval. A complete alternative quantization would,
however, require a separate analysis of the operator
domains, self-adjointness, and Hamiltonian dynamics.
Since the present work investigates the representation
based on Eq.~\eqref{eq:positive_momentum_measure} on its
positive sectors, we leave that alternative construction
for future work. Our sector restriction is therefore a
choice of quantization prescription, not a no-go theorem
for the remaining intervals.

The singular points themselves are excluded from the
regular domains because the reduced Poisson bracket and
$f(p_y)$ diverge there, while the measure in
Eq.~\eqref{eq:positive_momentum_measure} degenerates.

\paragraph{Inner product and operator representation.}

On the selected sectors, we introduce the inner product
\begin{equation}
    \langle\phi|\psi\rangle
    =
    \sum_{k\in\{1,3,5\}}
    \int_{I_k}
    \frac{dp_y}{f(p_y)}
    \,\phi_k^*(p_y)\psi_k(p_y),
    \label{eq:sector_restricted_inner_product}
\end{equation}
where $\phi_k$ and $\psi_k$ denote the wavefunction
components on the respective intervals. The corresponding
Hilbert space is the direct sum
\begin{equation}
    \mathcal H_+
    =
    L^2(I_1,D\,dp_y)
    \oplus
    L^2(I_3,D\,dp_y)
    \oplus
    L^2(I_5,D\,dp_y).
\end{equation}
The direct-sum structure retains the sector label, so
wavefunction components belonging to different intervals
are treated as distinct components of the state.

The momentum and position operators are represented
formally by
\begin{equation}
    (\hat p_y\psi)_k(p_y)=p_y\psi_k(p_y),
    \qquad
    (\hat y\psi)_k(p_y)
    =
    i\hbar f(p_y)\frac{d\psi_k}{dp_y}.
    \label{eq:momentum_operator_representation}
\end{equation}
On smooth test functions supported strictly inside the
selected intervals, this representation reproduces the
commutation relation in Eq.~\eqref{eq:deformed_commutator}.
Integration by parts shows that $\hat y$ is symmetric on
the corresponding compactly supported test-function domain.
This symmetry does not, by itself, establish self-adjointness
or determine the possible self-adjoint extensions. These
operator-domain questions are examined below for the
individual sectors.

\subsection{Operator Domains and the Natural Momentum Coordinate}
\label{subsec:operator_domains}

The momentum-dependent commutator derived above requires a
careful specification of operator domains. In particular, a
formal differential representation does not, by itself,
establish self-adjointness or guarantee that the position,
momentum, and Hamiltonian operators possess compatible
domains. The importance of these domain questions in
generalized uncertainty principle frameworks has been
emphasized in earlier work~\cite{Balasubramanian:2014pba}.
Here we examine them for the positive-measure momentum
sectors selected by the reduced symplectic structure.


Recall the reduced momentum map
\begin{equation}
    F(p_y)=p_y-\beta p_y^3+\beta'^2p_y^5,
\end{equation}
with
\begin{equation}
    D(p_y)=F'(p_y)
    =1-3\beta p_y^2+5\beta'^2p_y^4,
    \qquad
    f(p_y)=\frac{1}{D(p_y)}.
\end{equation}
The measure adopted in this work is
\begin{equation}
    d\mu(p_y)=\frac{dp_y}{f(p_y)}
    =D(p_y)\,dp_y.
\end{equation}
We restrict the quantum representation to the three regular
intervals on which this measure is positive,
\begin{equation}
    \mathcal P_+=I_1\cup I_3\cup I_5,
\end{equation}
where
\begin{equation}
    \begin{aligned}
    I_1&=(-\infty,-p_+),\\
    I_3&=(-p_-,p_-),\\
    I_5&=(p_+,\infty).
    \end{aligned}
\end{equation}
The intermediate intervals, on which $D(p_y)<0$, are not
included in the Hilbert-space construction considered here.
This restriction is specific to the measure and quantization
prescription adopted in this work.

On each selected regular sector, introduce the natural
momentum coordinate
\begin{equation}
    q=F(p_y).
    \label{eq:natural_momentum_coordinate}
\end{equation}
Since $D(p_y)>0$ on these sectors, the map is strictly
increasing on each of them and admits a local inverse
$p_y=p_y^{(k)}(q)$. Moreover,
\begin{equation}
    \frac{dq}{dp_y}=D(p_y)=\frac{1}{f(p_y)},
    \qquad
    \frac{dp_y}{f(p_y)}=dq.
\end{equation}
Consequently, the Hilbert space associated with a selected
sector can be written as
\begin{equation}
    \mathcal H_k=L^2(Q_k,dq),
\end{equation}
where $Q_k=F(I_k)$ denotes the image of that sector.

In the natural-coordinate representation, the formal operators
are
\begin{equation}
    \hat y=i\hbar\frac{d}{dq},
    \qquad
    \hat p_y=p_y^{(k)}(q).
    \label{eq:q_representation}
\end{equation}
For smooth test functions supported in the interior of $Q_k$,
these operators satisfy
\begin{equation}
    [\hat y,\hat p_y]\psi(q)
    =
    i\hbar\frac{dp_y^{(k)}}{dq}\psi(q)
    =
    i\hbar f\!\left(p_y^{(k)}(q)\right)\psi(q).
    \label{eq:commutator_q_representation}
\end{equation}
This identity holds on a common dense domain of test functions.
It does not, by itself, determine the self-adjoint realizations
of the observables or establish the domain properties of the
full Hamiltonian.

\paragraph{Central momentum sector.}

The central sector is
\begin{equation}
    I_3=(-p_-,p_-),
\end{equation}
where $p_->0$ is the smaller positive zero of $D(p_y)$.
Define
\begin{equation}
    q_-=F(p_-)>0.
\end{equation}
Since $F$ is odd and strictly increasing on $I_3$, its image is
\begin{equation}
    Q_3=(-q_-,q_-).
\end{equation}
On $L^2((-q_-,q_-),dq)$, the operator
$i\hbar\,d/dq$, initially defined on
$C_0^\infty((-q_-,q_-))$, admits self-adjoint extensions.
These extensions are characterized by the boundary conditions
\begin{equation}
    \psi(q_-)=e^{i\theta}\psi(-q_-),
    \qquad \theta\in[0,2\pi),
    \label{eq:position_boundary_condition}
\end{equation}
with domains
\begin{equation}
    \mathcal D(\hat y_\theta)=
    \left\{
    \psi\in H^1((-q_-,q_-)):
    \psi(q_-)=e^{i\theta}\psi(-q_-)
    \right\}.
    \label{eq:position_selfadjoint_domain}
\end{equation}
Thus, a self-adjoint realization of the position operator can
be chosen in the central-sector Hilbert space by specifying
the extension parameter $\theta$.

This result concerns the position operator alone. It does not
automatically establish that the momentum and Hamiltonian
operators have compatible domains or that the resulting
dynamics preserve the chosen operator domains. These questions
must be addressed before a complete quantum theory can be
claimed.

\paragraph{Outer momentum sectors.}

The two outer selected sectors are
\begin{equation}
    I_1=(-\infty,-p_+),
    \qquad
    I_5=(p_+,\infty),
\end{equation}
where $p_+>p_-$ is the larger positive zero of $D(p_y)$.
Their images under $q=F(p_y)$ are
\begin{equation}
    Q_1=(-\infty,-F(p_+)),
    \qquad
    Q_5=(F(p_+),\infty).
    \label{eq:outer_sector_images}
\end{equation}
For the illustrative parameter choice $\beta'=\beta/2$,
one finds $F(p_+)=0$, so that
\begin{equation}
    Q_1=(-\infty,0),
    \qquad
    Q_5=(0,\infty).
\end{equation}

On either individual half-line, the minimal operator
$i\hbar\,d/dq$ has unequal deficiency indices and therefore
does not admit a self-adjoint extension on that half-line
Hilbert space alone. Consequently, the formal position
operator cannot be interpreted as a self-adjoint observable
on either outer-sector Hilbert space independently.

A possible extension would be to investigate a larger
Hilbert space incorporating both outer sectors and to determine
whether a suitable self-adjoint realization exists on the
combined space. Such a construction would require explicit
boundary conditions and a specification of the domains of
the relevant observables. We do not assume that this extension
has been established here.

The selected intervals therefore define regular regions for
the sectorwise representation, but they do not, by themselves,
establish independent superselection sectors or a complete
global quantum theory. In particular, the compatibility of
the self-adjoint operator domains and the dynamics across the
critical momenta remains an open problem.

\subsection{Uncertainty Relation in the Bounded Sector \texorpdfstring{$\mathcal H_3$}{H3}}
\label{subsec:uncertainty_bounded_sector}

We first consider the bounded central momentum sector
\begin{equation}
    \mathcal H_3
    =
    L^2((-p_-,p_-),d\mu),
    \qquad
    d\mu=D(p_y)\,dp_y.
\end{equation}
Here $p_-$ denotes the smaller positive zero of $D(p_y)$.
Throughout this subsection, we consider normalized states for
which the relevant operator products, commutator expectation,
and uncertainties are well defined on a suitable common
domain.

For such states, the Robertson uncertainty relation gives
\begin{equation}
    \Delta y\,\Delta p_y
    \geq
    \frac{1}{2}
    \left|
    \langle[\hat y,\hat p_y]\rangle
    \right|.
\end{equation}
Using the reduced quantum algebra
\begin{equation}
    [\hat y,\hat p_y]
    =i\hbar f(\hat p_y),
    \label{eq:uncertainty_commutator}
\end{equation}
we obtain
\begin{equation}
    \label{exactuncertainty}
    \Delta y\,\Delta p_y
    \geq
    \frac{\hbar}{2}
    \left|\langle f(\hat p_y)\rangle\right|.
\end{equation}
Since $f(p_y)>0$ throughout the central sector, this becomes
\begin{equation}
    \label{exactGUP}
    \Delta y\,\Delta p_y
    \geq
    \frac{\hbar}{2}
    \left\langle
    \frac{1}{
    1-3\beta\hat p_y^2
    +5\beta'^2\hat p_y^4}
    \right\rangle.
\end{equation}
This GUP-type relation follows from quantizing the reduced
symplectic structure rather than postulating a phenomenological
deformation of the canonical commutator. Its interpretation
nevertheless depends on the existence of states in the
relevant common operator domain.

The expectation value in Eq.~\eqref{exactGUP} must be retained
in its exact form. In particular, the moment inequality
\begin{equation}
    \langle p_y^4\rangle
    \geq
    \langle p_y^2\rangle^2
\end{equation}
does not, by itself, imply
\begin{equation}
    \left\langle
    \frac{1}{
    1-3\beta p_y^2+5\beta'^2p_y^4}
    \right\rangle
    \geq
    \frac{1}{
    1-3\beta\langle p_y^2\rangle
    +5\beta'^2\langle p_y^2\rangle^2}.
\end{equation}
We therefore do not replace the exact expectation value by
a rational expression involving only momentum moments.

A useful lower bound follows from the support of the momentum
distribution. Define
\begin{equation}
    A(p_y)
    =
    1-3\beta p_y^2+5\beta'^2p_y^4,
    \qquad f(p_y)=\frac{1}{A(p_y)}.
\end{equation}
For $|p_y|<p_-$, one has
\begin{equation}
    0<A(p_y)\leq 1,
    \qquad f(p_y)\geq 1.
\end{equation}
It follows from Eq.~\eqref{exactGUP} that
\begin{equation}
    \label{eq:central_uncertainty_product}
    \Delta y\,\Delta p_y
    \geq \frac{\hbar}{2}.
\end{equation}

Consider normalized states satisfying
\begin{equation}
    \langle\hat p_y\rangle=0
\end{equation}
whose momentum distributions are supported strictly inside
$(-p_-,p_-)$. For these states,
\begin{equation}
    (\Delta p_y)^2
    =\langle\hat p_y^2\rangle
    <p_-^2,
    \qquad
    \Delta p_y<p_-.
\end{equation}
Combining this result with
Eq.~\eqref{eq:central_uncertainty_product} gives
\begin{equation}
    \label{positionbound}
    \Delta y>\frac{\hbar}{2p_-}.
\end{equation}

Equation~\eqref{positionbound} is a conditional lower bound
for the specified class of central-sector states. It requires
that the position uncertainty and the commutator expectation
be well defined on the relevant common domain. It does not
establish that the bound is optimal or attainable, nor does
it imply a universal minimal length for the full theory.
In particular, the existence of a self-adjoint extension of
the position operator on the bounded interval does not by
itself establish a compatible self-adjoint Hamiltonian. The
compatibility of the operator domains must be checked before
the bound is interpreted within a complete quantum dynamics.

For the illustrative parameter choice
\begin{equation}
    \beta=2\beta',
\end{equation}
the smaller critical momentum is
\begin{equation}
    p_-^2=\frac{2}{5\beta}.
\end{equation}
Equation~\eqref{positionbound} then gives
\begin{equation}
    \label{specificbound}
    \Delta y>
    \frac{\hbar}{2}\sqrt{\frac{5\beta}{2}}.
\end{equation}
The associated characteristic scale is
\begin{equation}
    \Delta y\sim\hbar\sqrt{\beta}.
\end{equation}
The coefficient in Eq.~\eqref{specificbound} characterizes
the derived lower bound; it should not be interpreted as
an attained minimum. Determining an optimal localization
scale would require a variational analysis over admissible
states together with a consistent specification of the
position and Hamiltonian domains.

\subsection{Uncertainty Structure in the Outer Sectors}
\label{subsec:uncertainty_outer_sectors}

We now consider the two outer positive-measure sectors,
\begin{equation}
    \mathcal H_1
    =
    L^2((-\infty,-p_+),d\mu),
    \qquad
    \mathcal H_5
    =
    L^2((p_+,\infty),d\mu),
\end{equation}
where
\begin{equation}
    d\mu=D(p_y)\,dp_y.
\end{equation}
These are the outer sectors of the positive-measure
representation adopted in this work.

Unlike the central sector, the outer sectors have unbounded
momentum spectra:
\begin{equation}
    p_y\in(-\infty,-p_+)
    \quad\text{or}\quad
    p_y\in(p_+,\infty).
\end{equation}
Consequently, these sectors do not impose a finite upper
bound on the magnitude of momentum. Moreover, the nonzero
threshold $|p_y|>p_+$ does not by itself imply a positive
lower bound on $\Delta p_y$, since a normalized momentum
distribution can be concentrated in an arbitrarily narrow
interval around a point strictly inside either sector.

The uncertainty analysis is constrained by the operator
domains. In the natural coordinate $q=F(p_y)$, the formal
position operator is
\begin{equation}
    \hat y=i\hbar\frac{d}{dq}.
\end{equation}
For the illustrative parameter choice $\beta'=\beta/2$,
the images of the outer sectors are
\begin{equation}
    Q_1=(-\infty,0),
    \qquad
    Q_5=(0,\infty).
\end{equation}
More generally, their endpoints are determined by
$\pm F(p_+)$.

On either individual half-line, the minimal derivative
operator has unequal deficiency indices and therefore admits
no self-adjoint extension on that half-line Hilbert space
alone. Thus, the formal differential expression cannot be
regarded as a self-adjoint position observable on either
individual outer sector without additional structure.

One possible extension would be to investigate a larger
Hilbert space incorporating both outer sectors and to determine
whether a self-adjoint realization can be defined on the
combined space. Such a construction would require explicit
operator domains and boundary conditions, together with
compatibility with the momentum and Hamiltonian operators.
We do not assume that this construction has been established
here.

Accordingly, although the deformed commutator holds formally
on suitable test-function domains, a complete Robertson
uncertainty analysis for the outer sectors requires a
consistent specification of the relevant self-adjoint
observables and their common domains. We therefore do not
infer an outer-sector localization bound from the momentum
threshold alone.

The sectorwise analysis consequently distinguishes the
bounded central momentum interval from the two unbounded
outer intervals. The central sector yields the conditional
bound in Eq.~\eqref{positionbound}, whereas the outer-sector
analysis does not establish an analogous bound. Neither
result, by itself, proves a universal minimal length, a
universal maximal momentum, or a complete quantum theory
across the critical momenta.

\subsection{Relation to Branched Quantization}
\label{subsec:relation_branched_quantization}

The construction developed above should be distinguished from
the branched quantization of Shapere and Wilczek~\cite{PhysRevLett.109.200402}.
The existence of a reduced classical formulation with a
single-valued Hamiltonian is not, by itself, the principal
novelty of the present work. Our aim is instead to investigate
the quantum representation and localization properties that
arise when the reduced, noncanonical phase-space structure is
quantized.

In the branched approach, the multivalued Hamiltonian and its
distinct branches are retained explicitly, and the quantum
description must account for the corresponding branch structure
and matching conditions. In the present approach, the Hamiltonian
is single-valued in the reduced variables, whereas the reduced
symplectic structure becomes singular at the critical momenta.
Quantization of this structure leads, on the regular sectors,
to a momentum-dependent commutation relation and a natural
momentum coordinate in which the position operator takes a
differential form.

The quantum consequences are not exhausted by the formal
deformation of the canonical commutator. The natural momentum
coordinate determines the sector Hilbert-space measures and
exposes differences between bounded and semi-infinite momentum
domains. In the bounded central sector, the deformed algebra
implies a conditional lower bound on position uncertainty for
states with vanishing mean momentum and support strictly inside
the central momentum interval. In contrast, on either individual
outer-sector Hilbert space, the formal position operator does
not admit a self-adjoint extension. Thus, the operator-domain
analysis reveals a distinction between the central and outer
sectors that is not apparent from the commutation relation alone.

The main result of this analysis is a characterization of the
operator-domain structure associated with the selected
positive-measure representation of the reduced phase-space
algebra. In particular, the bounded central sector admits
self-adjoint extensions of the position operator and yields
a conditional localization bound, whereas the position
operator on either individual outer half-line has no
self-adjoint extension. These results do not establish a
complete quantum dynamics or resolve the equivalence with
branch-matching quantization. Rather, they identify specific
operator-theoretic questions that must be settled to determine
whether the reduced formulation defines a consistent global
quantum theory.

These results characterize the quantum structure of the
sectorwise reduced formulation. They do not establish that
the sectors are independent superselection sectors, nor
that the reduced formulation is equivalent to, or physically
distinct from, the branched quantization of Shapere and
Wilczek. Such conclusions require a more complete comparison
of the Hilbert spaces, operator domains, self-adjoint
realizations, boundary or matching conditions, spectra, and
physical observables. The degeneracy of the reduced symplectic
structure at the critical momenta also means that a prescription
for the global dynamics across these points cannot be inferred
from the regular-sector construction alone.

\begin{table*}[t]
\centering
\caption{Comparison between the branched quantization of
Shapere and Wilczek~\cite{PhysRevLett.109.200402} and the sectorwise reduced
quantization investigated in the present work. The table
summarizes the formulations and results considered here;
it does not establish their quantum equivalence or
inequivalence.}
\label{tab:comparison}
\renewcommand{\arraystretch}{1.25}
\setlength{\tabcolsep}{6pt}
\begin{tabular}{p{0.18\textwidth} p{0.36\textwidth} p{0.36\textwidth}}
\hline
\textbf{Aspect}
&
\textbf{Branched quantization}
&
\textbf{Present work}
\\
\hline

Classical formulation
&
Retains the explicit branch structure of the Hamiltonian.
&
Uses a single-valued reduced Hamiltonian with a
momentum-dependent, singular reduced symplectic structure.
\\[3pt]

Quantum description
&
Treats the quantum dynamics in the presence of the
classical branch structure.
&
Quantizes the reduced phase-space algebra on selected
regular momentum sectors.
\\[3pt]

Critical points
&
Arise in connection with the branching of the classical
Hamiltonian.
&
Correspond to zeros of the reduced symplectic Jacobian
$D(p_y)$, where the reduced symplectic structure becomes
degenerate.
\\[3pt]

Operator analysis
&
Requires a consistent quantum treatment of the branch
structure and its associated operator domains.
&
Examines the natural momentum coordinate and the
self-adjointness properties of the position operator
on the selected sectors.
\\[3pt]

Localization
&
Depends on the quantum representation, its operator
domains, and its admissible states.
&
Derives a conditional lower bound on position uncertainty
in the bounded central sector for states with
$\langle p_y\rangle=0$ and support strictly inside
$(-p_-,p_-)$.
\\[3pt]

Global quantum status
&
The relevant quantum construction must specify the
domains and treatment of the branch structure.
&
Global domain compatibility, sector matching, and
equivalence with branched quantization remain unresolved.
\\
\hline
\end{tabular}
\end{table*}

Table~\ref{tab:comparison} summarizes the distinction between
the explicit-branch formulation and the sectorwise reduced
representation studied here. The principal result of the
present analysis is the characterization of the position
operator on the selected regular momentum sectors. The
bounded central sector admits self-adjoint extensions of
the position operator and yields a conditional lower bound
on position uncertainty, whereas the formal position
operator on either individual outer half-line admits no
self-adjoint extension on that half-line Hilbert space
alone. These results characterize the operator-domain
structure of the chosen representation but do not establish
a complete global quantum dynamics. In particular, whether
the reduced formulation is unitarily equivalent to, or
physically distinct from, branched quantization remains
an open question requiring a comparison of the corresponding
Hilbert spaces, operator domains, and physical observables.

\subsection{Formal Spectral Comparison in a Harmonic Potential}
\label{subsec:spectral_comparison}

To clarify the relation between the reduced and branch-resolved
descriptions at the operator level, we consider a harmonic
confining potential. Our aim is to compare their formal
differential expressions and identify the additional domain
information required for a meaningful spectral comparison.

We take
\begin{equation}
    V(y)=\frac12m\omega^2y^2.
\end{equation}
Using the reduced Hamiltonian derived in Eq.~\eqref{c}, the
kinetic contribution is
\begin{equation}
    T(p_y)=
    \frac{p_y^2}{2m}
    -\frac{3\beta}{4m}p_y^4
    +\frac{5\beta'^2}{6m}p_y^6,
\end{equation}
and the canonical momentum is related to the mechanical
momentum through
\begin{equation}
    P=F(p_y)
    =p_y-\beta p_y^3+\beta'^2p_y^5.
\end{equation}

On a regular classical branch $I_j$, let $p_y^{(j)}(P)$
denote the corresponding inverse of the momentum map, defined
on the image $F(I_j)$. The branchwise Hamiltonian is
\begin{equation}
    H_j(y,P)=
    T\!\left(p_y^{(j)}(P)\right)
    +\frac12m\omega^2y^2.
\end{equation}
In the canonical momentum representation, the formal
position operator is $\hat y=i\hbar\,d/dP$. The corresponding
formal stationary Schr\"odinger equation is
\begin{equation}
    \left[
    -\frac{m\omega^2\hbar^2}{2}\frac{d^2}{dP^2}
    +T\!\left(p_y^{(j)}(P)\right)
    \right]\psi_j(P)
    =E\psi_j(P).
    \label{eq:branched_spectral_equation}
\end{equation}
This equation specifies a local differential expression on
the relevant branch image. Defining a quantum spectral
problem additionally requires a Hilbert space and an
appropriate operator domain.

In the reduced formulation, the natural coordinate
$q=F(p_y)$ coincides with the canonical momentum $P$.
Consequently, the reduced and branchwise formulations have
the same local differential expression on any corresponding
regular sector. This establishes a formal local identity of
the spectral operators, but does not establish equality of
their spectra. Spectral equivalence also depends on the
Hilbert spaces, operator domains, and boundary conditions.

\paragraph{Critical points and the retained quantum sectors.}

For the illustrative parameter choice
\begin{equation}
    \beta'=\frac{\beta}{2},
\end{equation}
the positive critical mechanical momenta satisfy
\begin{equation}
    p_-^2=\frac{2}{5\beta},
    \qquad
    p_+^2=\frac{2}{\beta}.
\end{equation}
Their images under the momentum map are
\begin{equation}
    Q_-=F(p_-)
    =\frac{16}{25}\sqrt{\frac{2}{5\beta}},
    \qquad
    F(p_+)=0.
\end{equation}
Since $F$ is odd, the corresponding negative critical
momenta have images $-Q_-$ and $0$, respectively.

The positive-measure representation adopted in this work
retains the three regular sectors $I_1$, $I_3$, and $I_5$,
on which
\begin{equation}
    D(p_y)=\frac{dF}{dp_y}
    =1-3\beta p_y^2+5\beta'^2p_y^4>0.
\end{equation}
In the natural coordinate $q=F(p_y)$, the images of these
sectors are
\begin{equation}
    Q_1=(-\infty,0),\qquad
    Q_3=(-Q_-,Q_-),\qquad
    Q_5=(0,\infty).
\end{equation}
Thus, the central sector corresponds to a finite interval,
whereas each outer sector corresponds to a half-line. The
images overlap in the natural momentum coordinate, so the
sector label remains necessary to distinguish the
corresponding reduced representations.

The inner product in the reduced representation is defined
using
\begin{equation}
    d\mu=D(p_y)\,dp_y=dq
\end{equation}
on each retained sector. The density $D(p_y)$ changes sign
on the full mechanical-momentum line, so this expression
does not define a positive measure on the full line. The
restriction to $I_1$, $I_3$, and $I_5$ is therefore part of
the positive-measure representation developed here. It does
not, by itself, establish that the other classical intervals
cannot be quantized using a different positive-inner-product
construction.

\paragraph{Scope of the spectral comparison.}

The local identity of the differential expressions in
Eq.~\eqref{eq:branched_spectral_equation} shows that a distinct
quantum spectrum cannot be inferred merely from the
single-valued form of the reduced Hamiltonian or from the
momentum-dependent commutator. A meaningful spectral
comparison requires a well-defined self-adjoint Hamiltonian
in each formulation, together with a consistent specification
of the associated Hilbert spaces and operator domains.

In the central reduced sector, the formal position operator
on the finite natural-coordinate interval admits
self-adjoint extensions. This does not, by itself, establish
self-adjointness of the full harmonic-oscillator Hamiltonian:
the domain of the Hamiltonian, including the operator
associated with the quadratic potential, must also be
specified and checked for consistency. On either individual
outer half-line, the formal derivative operator does not
admit a self-adjoint extension on that half-line Hilbert
space alone. A complete spectral analysis of the outer
sectors must therefore address the operator domains and
determine whether an enlarged Hilbert space or additional
sector coupling is required.

A branch-resolved comparison likewise requires a justified
prescription for the operator domains and any matching
conditions at the critical points. Such conditions cannot
be inferred solely from the existence of multiple classical
branches. We therefore do not claim a new spectral prediction
or quantum inequivalence at this stage. Determining whether
a consistent completion of the reduced formulation is
unitarily equivalent to a branch-resolved quantization
remains an open problem.

\section{Conclusion}

In this work, we have investigated the quantum consequences of the reduced phase-space geometry associated with classical systems exhibiting spontaneous breaking of continuous time-translation symmetry. In these systems, the non-convex velocity dependence of the Lagrangian renders the canonical momentum--velocity relation non-invertible, leading to a branched Hamiltonian description \cite{PhysRevLett.109.200402}. The reduced formulation considered here provides an alternative description in terms of a single-valued Hamiltonian, with the critical structure of the classical dynamics encoded in a singular, momentum-dependent symplectic geometry.

The contribution of the present work is to investigate the quantum kinematic consequences of this reduced geometry, rather than to introduce the classical reduction itself. We construct a momentum-space representation based on the reduced symplectic measure and examine how the classical critical structure constrains the admissible quantum momentum sectors. In particular, the critical mechanical momenta appear as singular points of the momentum-dependent commutation relation. In the stabilized model considered here, the outer critical momenta also coincide with the finite-momentum ground-state configurations associated with persistent classical motion. This correspondence connects the critical structure of the classical dynamics to the resulting quantum kinematics.

Using the positive measure induced by the reduced symplectic structure, we identify three admissible regular momentum sectors: a bounded central sector and two unbounded outer sectors. This sector structure characterizes the positive-inner-product representation adopted here. Within the central sector, the deformed commutation relation yields a nonzero lower bound on position uncertainty for states with vanishing mean momentum, subject to the operator-domain assumptions specified in the text. This bound arises from the reduced phase-space geometry rather than from an independently postulated generalized uncertainty principle. However, its attainability by a normalizable physical state has not been established, and it should therefore not be interpreted as a universal minimal length. Similarly, the unbounded momentum support of the outer sectors does not imply a universal maximal momentum.

These results provide a complementary perspective on the branched quantization developed by Shapere and Wilczek \cite{PhysRevLett.109.160402,PhysRevLett.109.200402}. Whereas the branched description retains the distinct Hamiltonian branches and requires a prescription for relating them, the present approach begins with the reduced phase-space geometry and examines how its singular structure is reflected in the quantum algebra, inner product, and admissible momentum sectors. Our formal spectral comparison shows that the reduced and branchwise descriptions have the same local differential expression when written in the natural momentum coordinate. This local identity does not, however, establish equivalence of their quantum spectra. Determining whether the two formulations are unitarily equivalent or define inequivalent quantum theories requires a comparison of their Hilbert spaces, self-adjoint Hamiltonians, operator domains, and physical observables.

The \(k\)-essence example discussed in Sec.~II~C suggests a possible extension of the present framework to field-theoretic systems. For the polynomial scalar-field model considered there, the Jacobian of the homogeneous velocity--momentum map vanishes at finite scalar velocities for an appropriate range of kinetic couplings. Equivalently, the same degeneracy condition appears in the scalar kinetic coefficient entering the sound-speed expression. This provides a structural analogy with the critical points of the mechanical Legendre map studied in this work, but does not establish that the coupled gravity--scalar system possesses the same reduced phase-space structure. An important direction for future work is to determine whether this degeneracy survives the imposition of the gravitational constraints and the reduction to the physical phase space. If it does, one can investigate whether the resulting reduced symplectic geometry constrains the admissible quantum representations and the evolution of scalar-field states near the critical velocities. In particular, it remains to be established whether distinct quantum sectors emerge and whether consistent quantum evolution across the degeneracy requires additional boundary conditions or a suitable completion. The vanishing of the Legendre-map Jacobian alone does not establish these consequences; they depend on the reduced symplectic structure of the physical degrees of freedom. We leave this investigation for future work.

Several questions therefore remain open. A complete specification of the quantum theory requires a systematic analysis of the operator domains and boundary conditions, including the self-adjointness of the full Hamiltonian in each retained sector. The relation between the reduced and branched quantizations must also be established, and the physical attainability of the localization bound requires further investigation. Applying the framework to explicit regularized models may clarify which features of the resulting quantum structure are generic and which depend on the particular stabilized dynamics considered here.

In summary, this work provides a quantum-kinematic framework for investigating spontaneous time-translation-symmetry breaking through the singular geometry of a reduced phase space. By relating the classical critical momenta to singularities of the quantum algebra, identifying the positive-inner-product momentum sectors, and deriving a conditional localization bound, we show how information associated with classically branched dynamics can remain relevant in a quantum formulation with a single-valued Hamiltonian. Establishing the self-adjoint quantum dynamics, determining the relationship to branched quantization, and investigating whether analogous quantum structures emerge near Legendre-map folds in constrained scalar--gravity systems constitute important directions for future work.

\section{Acknowledgements}

This work was supported by the RPF Postdoctoral Fellowship
Program. I thank the members of the Department of Physics,
Kyushu University, Japan, for their gracious hospitality during
my visit toward the completion of this work. I am grateful to
Prof.~Kazuhiro Yamamoto and Prof.~Akira Matsumura for fruitful
discussions. I also thank Prof.~Frederick G.~Scholtz and
Prof.~Douglas Singleton for useful correspondence and discussions.

\bibliographystyle{apsrev4-1}

\bibliography{gw_phases}

\end{document}